\documentclass{aastex63}
\usepackage{graphicx,natbib}
\usepackage[percent]{overpic}
\usepackage{graphicx,subfigure,amsmath}
\usepackage{hyperref}
\usepackage[all]{nowidow}
\newcommand\gsim{\,\lower3pt\hbox{$\sim$}\llap{\raise2pt\hbox{$>$}}\,}
\newcommand\lsim{\,\lower3pt\hbox{$\sim$}\llap{\raise2pt\hbox{$<$}}\,}

\shorttitle{Simulation of AR11158 eruption}
\shortauthors{Fan et al.}

\begin{document}

\title{A data-driven MHD simulation of the 2011-02-15 coronal mass ejection from Active Region NOAA 11158}

\correspondingauthor{Yuhong Fan}
\email{yfan@ucar.edu}

\author{Yuhong Fan}
\affil{High Altitude Observatory, National Center for Atmospheric Research, \\
3080 Center Green Drive, Boulder, CO 80301, USA}

\author{Maria D. Kazachenko}
\affiliation{Department of Astrophysical and Planetary Sciences, University of Colorado Boulder, Boulder, CO, USA}
\affiliation{National Solar Observatory, University of Colorado Boulder, Boulder, CO, USA}

\author{Andrey N. Afanasyev}
\altaffiliation{DKIST Ambassador}
\affiliation{Laboratory for Atmospheric and Space Physics, University of Colorado Boulder, CO, USA}
\affiliation{National Solar Observatory, University of Colorado Boulder, Boulder, CO, USA}
\affiliation{Institute of Solar-Terrestrial Physics of SB RAS, Irkutsk, Russia}

\author{George H. Fisher}
\affiliation{Space Sciences Laboratory, University of California Berkeley, Berkeley, CA, USA}

\begin{abstract}
We present a boundary data-driven magneto-hydrodynamic (MHD) simulation of the 2011-02-15 coronal mass ejection (CME) event of Active Region (AR) NOAA 11158. The simulation is driven at the lower boundary with an electric field derived from the normal magnetic field and the vertical electric current measured from the Solar Dynamics Observatory (SDO) Helioseismic Magnetic Imager (HMI) vector magnetograms. The simulation shows the build up of a pre-eruption coronal magnetic field that is close to the nonlinear force-free field (NLFFF) extrapolation, and it subsequently develops multiple eruptions. The sheared/twisted field lines of the pre-eruption magnetic field show qualitative agreement with the brightening loops in the SDO Atmospheric Imaging Assembly (AIA) hot passband images. We find that the eruption is initiated by the tether-cutting reconnection in a highly sheared field above the central polarity inversion line (PIL) and a magnetic flux rope with dipped field lines forms during the eruption. The modeled erupting magnetic field evolves to develop a complex structure containing two distinct flux ropes and produces an outgoing double-shell feature consistent with the Solar TErrestrial RElations Observatory B / Extreme UltraViolet Imager (STEREO-B/EUVI) observation of the CME. The foot points of the erupting field lines are found to correspond well with the dimming regions seen in the SDO/AIA observation of the event. These agreements suggest that the derived electric field is a promising way to drive MHD simulations to establish the realistic pre-eruption coronal field based on the observed vertical electric current and model its subsequent dynamic eruption.

\end{abstract}


\section{Introduction}
Large scale solar eruptions such as solar flares and coronal mass ejections are
major drivers of space weather near Earth \citep[e.g. review by][]{Temmer:2021}.
These eruptive phenomena are all
manifestations of the explosive release of magnetic energy stored in the non-potential,
current carrying coronal magnetic fields built up over time due to magnetic flux
emergence from the interior and shear and twisting motions at the interior
footpoints of the coronal field lines \citep[e.g.][]{Forbes:etal:2006,
Green:etal:2018,Patsourakos:etal:2020}. Thus determining the realistic 3D
coronal magnetic field evolution of the solar eruptive events is essential to
understanding the physical mechanisms that have led to the eruptions and also
advancing the capability of predicting their space weather impact.
In recent years, simulations using observed data for constructing the initial
and boundary conditions, called ``data-constrained" and ``data-driven"
simulations, have undergone significant development and been applied
to study real solar eruptive events with
complex magnetic structures \citep[see e.g. recent review by][]{Jiang:etal:2022}.
For example, these approaches have been extensively explored to model the magnetic
field evolution of the X2.2 flare and the associated halo CME on 2011–02-15
from Active Region (AR) NOAA 11158 \citep[e.g.][]{Cheung:DeRosa:2012,Inoue:etal:2014,
Inoue:etal:2015,Hayashi:etal:2018,Hoeksema:etal:2020,Afanasev:etal:2023},
which is a well observed quadrupolar, $\delta$-sunspot active region.

\citet{Cheung:DeRosa:2012} and \citet{Hoeksema:etal:2020} have developed
a frame work of boundary data-driven magneto-frictional (MF) modeling of the
force-free coronal magnetic field evolution of AR 11158, using an
electric field inverted from the HMI vector magnetograms
\citep{Kazachenko:etal:2014,Fisher:etal:2020} as the lower boundary driving
condition. They are able to model the quasi-static build up of the active region
magnetic field on realistic time scales, but cannot model the dynamic eruption
phase with the MF approach.  
\citet{Inoue:etal:2014} and \citet{Inoue:etal:2015} used a NLFFF extrapolation
at about 2 hours before the onset of the eruptive flare of AR 11158 as the initial
state and modeled the subsequent dynamic eruption with the zero-$\beta$
MHD simulation.  It is found that the twisted field of the initial NLFFF is stable
and that strongly twisted field lines are formed via the tether-cutting reconnection,
which is responsible for the onset of the eruption.
\citet{Afanasev:etal:2023} used a hybrid approach where the build up of the
active region is modeled with the data-driven MF simulation and a snapshot
from the MF simulation at a time about 1.5 hours before the onset of the eruption
is used as the initial magnetic field configuration for the MHD simulation to
model the subsequent dynamic eruption. It is found that the initial magnetic field
is already out of equilibrium and erupts immediately while at the same time
also going through an initial relaxation, so it is difficult for
the simulation to assess the initiation mechanism of the eruption.
Thus a data-driven MHD simulation that models both the quasi-static build up and
transition to dynamic eruption is needed to examine the initiation mechanisms.
However it is still not feasible for such MHD simulations to model the long
build up phase of the active region on realistic time scales.
\citet{Hayashi:etal:2018} has developed a data-driven MHD simulation of AR 11158
driven with a lower boundary electric field inverted from 
the temporal evolution of the three components of the vector magnetic field from
the HMI vector magnetograms.  Applying this electric field on an accelerated time
scale, their MHD simulation was able to reproduce the observed temporal evolution of
the (smoothed) photospheric magnetic field at the lower boundary and build up
a pre-eruption coronal magnetic field with sufficient free magnetic
energy to drive the X-class flare, but did not result in the release of the
magnetic energy and the development of the eruption.

In this work, we use a new electric field derived from the
observed normal magnetic field $B_r$ and vertical electric current $J_r$ evolution
from the SDO/HMI vector magnetograms for the lower boundary driving of
an MHD simulation
of the eruptive flare and CME developed from AR 11158 on Feb. 15, 2011.
The preliminary results of the simulation were reported in a NASA Living With a
Star focused science team joint paper \citep[section 4 of][]{Linton:etal:2023}.
Here we present a more detailed description of the simulation and the results,
and expand on the analysis of the erupting magnetic field and comparison with
the observations by the STEREO-B EUVI and the SDO/AIA.
We found the build up of a pre-eruption magnetic field that is close to the NLFFF
extrapolation and the onset of the subsequent eruption that reproduces some of the
observed features of the CME by the STEREO-B EUVI and the SDO/AIA.
The paper is organized as follows. In \S \ref{sec:model} we describe the
setup of the MHD simulation and the formulation of the lower boundary driving
electric field.  In \S \ref{sec:result} we present the simulation results and
comparison with the observations. The conclusions and a discussion are given
in \S \ref{sec:summary}

\section{Description of the data-driven simulation}
\label{sec:model}
An earlier description of the setup of the simulation was given in section
4 of \citet{Linton:etal:2023} which reported preliminary results of the simulation.
Here for the completeness of this paper, we provide a more detailed
description of the simulation setup and the lower boundary driving electric field used.

\subsection{The numerical model}
The data-driven simulation is carried out using the ``Magnetic Flux Eruption'' (MFE)
code which solves the set of semi-relativistic MHD equations as described
in \citet[][hereafter F17]{Fan:2017}.
The readers are referred to that paper for a description of the equations
solved and the numerical methods. Here we only describe the changes made
specifically for the setup of the current simulation.

As described in F17, the momentum equation includes the Boris correction with a
reduced speed of light to limit the Lorentz force, and hence relax the stringent Courant
condition on the numerical time-stepping due to the extremely high Alfv\'en speed present
in the simulation domain which contains a strong active region.
For this simulation, we have used a reduced speed of light of
about $8000$ km/s, which remains significantly higher than the peak plasma flow speed and
the the sound speed reached in the simulation.
For the thermodynamics, we assume an ideal gas of fully
ionized hydrogen, with $\gamma=5/3$, and solve the internal energy equation taking into
account the following non-adiabatic effects that include the field aligned thermal
conduction, optically thin radiative cooling, and coronal heating.
However, we no longer include an empirical coronal heating \citep[eq. (14) in][]{Fan:2017}
for the heating term in the internal energy equation
\citep[term H in eq. (5) of][]{Fan:2017},
but only include the resistive and viscous dissipation due to the numerical diffusion
\citep[as described on p.3 of][]{Fan:2017}.
For this simulation, we add to the lower boundary a random
electric field (as described below in \S \ref{sec:efield}) representing the effect of turbulent convection
that drives field line braiding, and the resultant (numerical) resistive and viscous
heating provides the heating of the corona.
This heating varies spatially and temporally self-consistently with the
formation of the strong current layers in the 3D magnetic field.

\begin{figure}[htb!]
\centering
\includegraphics[width=0.5\textwidth]{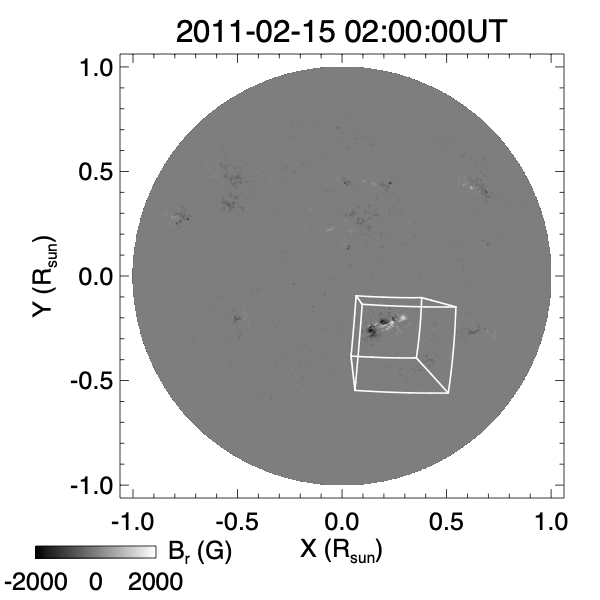}
\caption{The simulation domain against the solar disk as viewed from the Earth
perspective on 2011-02-15 at 02:00:00 UT. The gray scale image shows the
full-disk SDO/HMI magnetogram of the normal magnetic field $B_r$.}
\label{fig:domain}
\end{figure}

The simulation is carried out in a spherical wedge domain whose lower boundary is
centered on and tracked the active region. Figure \ref{fig:domain} shows the simulation
domain against the solar disk as viewed from the Earth perspective on
2011-02-15 at 02:00:00 UT.
It has a latitudinal width of 17.2 degrees,
a longitudinal width of 18.8 degrees, and a radial range of 1 to 1.43 solar
radius. The simulation domain is resolved with a grid of
$360 (r) \times 288 (\theta) \times 315 (\phi)$.
The grid is stretched in the radial dimension (with the radial grid size
$\Delta r$ increasing outward as a geometric series) and is uniform in
the two horizontal dimensions.  The peak radial resolution is 300 km
at the bottom, and the horizontal resolution is 724 km at the bottom.

\subsection{Formulation of the lower boundary electric field}
\label{sec:efield}
For the lower boundary magnetic flux transport driving conditions,
we impose a time dependent horizontal electric field ${\bf E}_h$ that consists
of three components:
\begin{equation}
{\bf E}_h = {\bf E}_h^P + {\bf E}_h^{\rm Tw} + {\bf E}_h^{\rm random}.
\label{eq:tot_efield}
\end{equation}
The first component ${\bf E}_h^P$ corresponds to the horizontal component of
the so-called ``PTD" (Poloidal-Toroidal-Decomposition) electric field derived in
\citet[][see the horizontal component of eq. (8) in that paper]{Fisher:etal:2020}:
\begin{equation}
c{\bf E}_h^P = - \nabla \times (\dot{P} \hat{\bf r})
\label{eq:P_efield}
\end{equation}
where $\dot{P}$ is computed from the observed normal magnetic field ($B_r$)
evolution on the photosphere by solving the 2D Poisson equation \citep[eq. (9)
in][]{Fisher:etal:2020}:
\begin{equation}
\nabla_h^2 \dot{P} = - {\dot{B}}_r,
\label{eq:P}
\end{equation}
due to Faraday's law. Imposing ${\bf E}_h^P$ reproduces the observed
photospheric normal magnetic field evolution on the lower boundary.

The second term ${\bf E}_h^{\rm Tw}$ on the right hand side
of equation (\ref{eq:tot_efield}) is the ``twisting" electric field.
It is given as the horizontal gradient of
a potential:
\begin{equation}
c {\bf E}_h^{\rm Tw} = - \nabla_h {\psi}^{\rm Tw}
\label{eq:twist_efield}
\end{equation}
such that it does not alter the observed $B_r$ evolution
(already reproduced by imposing the ${\bf E}_h^P$ component),
and we assume that it corresponds to a vertical transport of a
horizontal magnetic field into the domain, i.e.
\begin{equation}
- \nabla_h {\psi}^{\rm Tw} = -v_0 {\hat r} \times {\bf {\cal B}}_h
\label{eq:twist_potential}
\end{equation}
where $v_0$ is the vertical transport speed, which we assume to be
a constant, and ${\bf {\cal B}}_h$ is
a horizontal magnetic field, which we note is {\it not} the same as the
observed horizontal field
and it can be shown from equation (\ref{eq:twist_potential})
that $\nabla_h \cdot {\bf {\cal B}}_h = 0$ (which condition is
generally not satisfied by the observed horizontal magnetic field).
Further taking the horizontal divergence of
equation (\ref{eq:twist_potential}) yields:
\begin{equation}
\nabla_h^2 {\psi}^{\rm Tw} = - v_0 (\nabla_h \times {\bf {\cal B}}_h)_r .
\end{equation}
We let $(\nabla_h \times {\bf {\cal B}}_h)_r$ be equal to
$J_r = (\nabla_h \times {\bf B}^{\rm obs}_h)_r$, which is the
vertical electric current derived from the observed horizontal
magnetic field ${\bf B}^{\rm obs}_h$ in the photospheric vector magnetograms.
Thus we have
\begin{equation}
\nabla_h^2 {\psi}^{\rm Tw} = - v_0 J_r .
\label{eq:psi}
\end{equation}
We compute the potential ${\psi}^{\rm Tw}$ by solving the 2D Poisson
equation (\ref{eq:psi}) given the measured vertical electric current at
the photosphere $J_r$ from the HMI
vector magnetograms and specifying $v_0$, which is an {\it ad hoc}
parameter we can adjust. For the simulation presented here we have used
$v_0  = 2.5 \, {\rm km/s}$, which is small enough to ensure a quasi-static
evolution for the build-up phase but high enough to compete with numerical
diffusion to produce the eruptive behavior close to that observed.
Once ${\psi}^{\rm Tw}$ is determined, the
twisting electric field ${\bf E}_h^{\rm Tw}$ is given by equation
(\ref{eq:twist_efield}). Imposing ${\bf E}_h^{\rm Tw}$ at the lower
boundary corresponds to transporting a (divergence-free) horizontal
magnetic field (${\bf {\cal B}}_h$) with the observed vertical electric current
into the domain, without changing the normal
magnetic field evolution at the lower boundary.
It effectively transports twist into the corona based on the
observed vertical electric current.

\begin{figure}[htb!]
\centering
\includegraphics[width=0.24\textwidth]{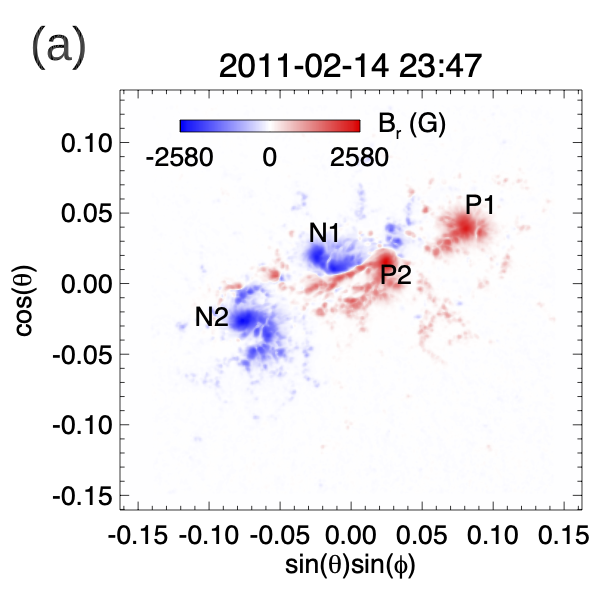}
\includegraphics[width=0.24\textwidth]{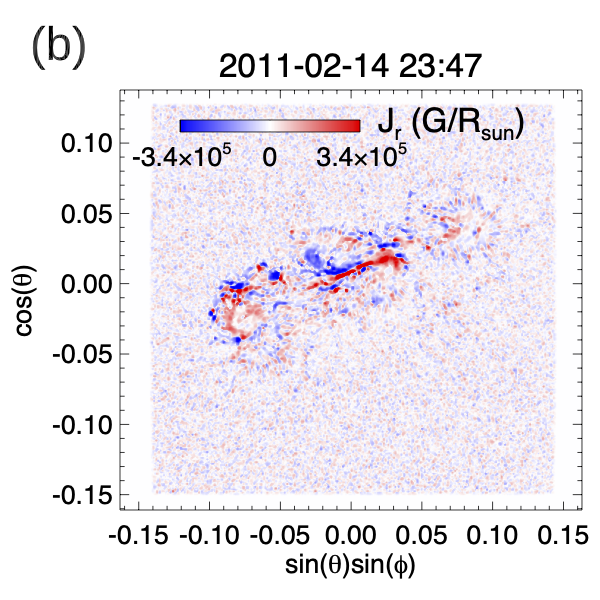}
\includegraphics[width=0.24\textwidth]{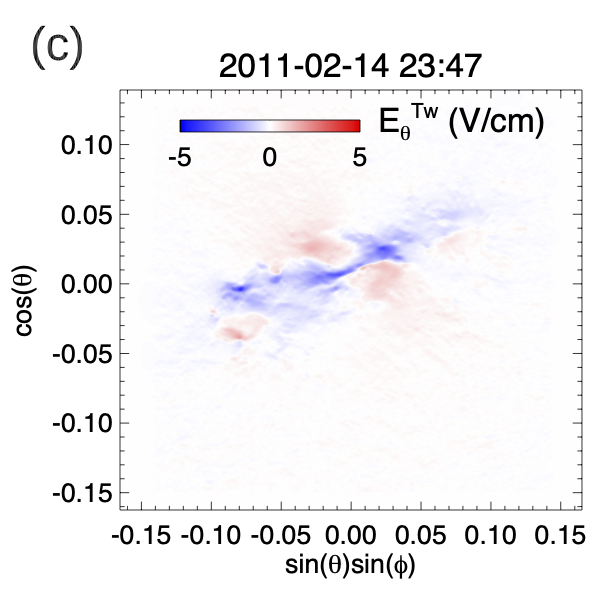}
\includegraphics[width=0.24\textwidth]{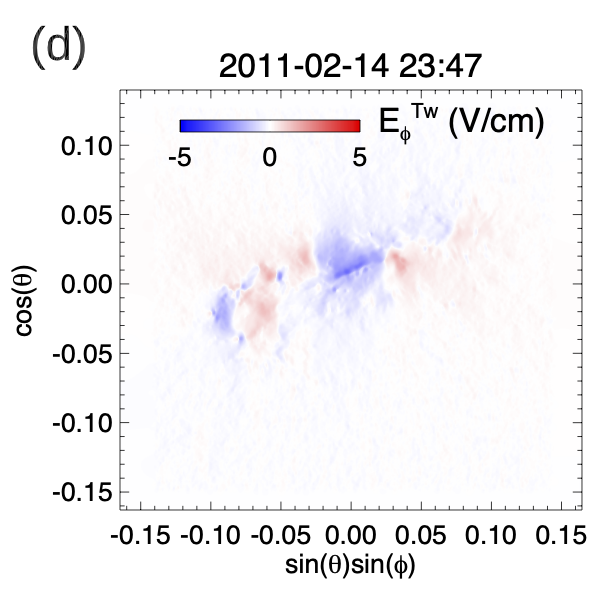} \\
\includegraphics[width=0.3\textwidth]{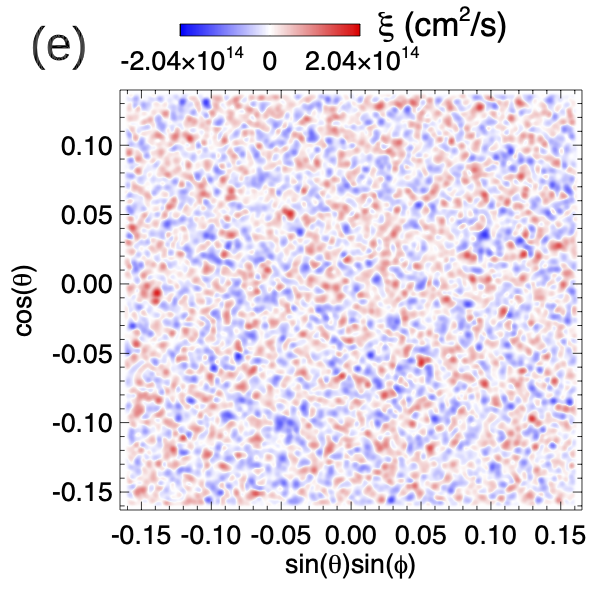}
\includegraphics[width=0.24\textwidth]{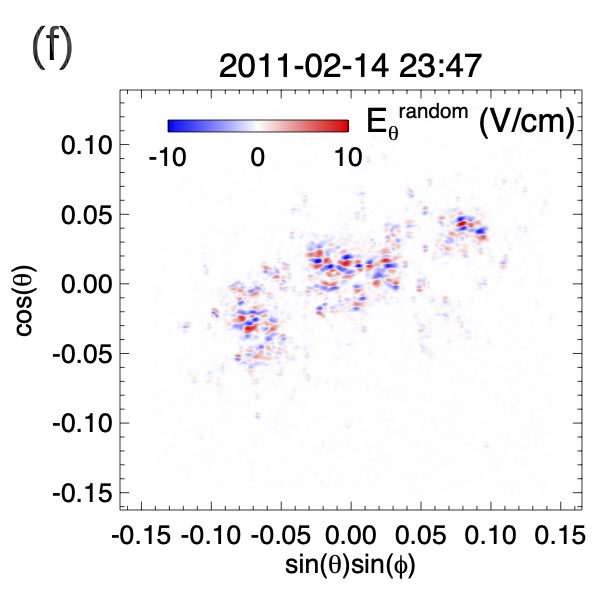}
\includegraphics[width=0.24\textwidth]{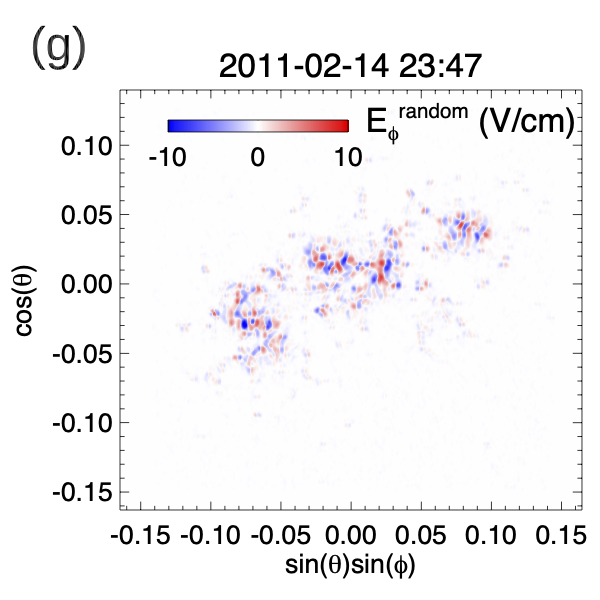}
\caption{Top row shows example snapshots of the oberved vertical
magnetic field $B_r$ (a), vertical electric current density $J_r$ (b),
and the horizontal twisting electric field components
$E_{\theta}^{\rm Tw}$ (c), and $E_{\phi}^{\rm Tw}$ (d), at the lower
boundary at 23:47 UT on 2011-02-14, about 1.9 hour before the onset
of the observed X-class flare.
The bottom row shows a snap shot of the $\xi$ field (e) for generating
the random electric field ${\bf{E}}_{h}^{\rm{random}}$ in
equation (\ref{eq:random_efield}), and the resulting random electric
field components $E_{\theta}^{\rm random}$ (f), and $E_{\phi}^{\rm random}$
(g) produced at the lower boundary when the $\xi$ field is used with
the $B_r$ given in panel (a).}
\label{fig:twistemf}
\end{figure}

Figure \ref{fig:twistemf} shows example snapshots of
the observed normal magnetic field $B_r$, the vertical
electric current density $J_r$, and the
horizontal twisting electric field components $E_{\theta}^{\rm Tw}$ and
$E_{\phi}^{\rm Tw}$ derived based on $J_r$ (from
eqs. [\ref{eq:twist_efield}] and [\ref{eq:psi}]),
at the time of about 1.9 hour before the onset of the observed
X-class flare.
From panels (a) and (b) we can see that significant $J_r$ which tends
to be of the same sign as $B_r$ is present in the polarity concentrations
P2 and N1 near the central polarity inversion line (PIL).
On the other hand, in the polarity concentration N2, we find
$J_r$ of predominantly opposite sign of $B_r$.
As a result, the derived twisting electric field (panels (c) and (d))
transports positive twist into the corona at the central P2, N1 flux
concentrations and negative twist at the N2 flux concentration.
It can be seen from panels (c) and (d) that $E_{\theta}^{\rm Tw}$,
and $E_{\phi}^{\rm Tw}$ transport a concentrated horizontal, shear
magnetic field component into the corona at the central PIL.
They also effectively drive a clockwise rotation of the P2 and N1
flux concentrations and a counter-clockwise rotation of the N2 flux
concentration.

We note that \citet{Cheung:DeRosa:2012} have also used a
horizontal electric field to drive twist into the corona in their
magneto-frictional modeling of the evolution of AR 11158.
In that case, effectively a uniform rotation of all the
polarity concentrations is applied to drive the same sign
of twist into the corona.  Here with the twisting electric field,
the twist injection varies spatially based on the distribution of
the observed vertical electric current.

The last term on the right hand side of equation (\ref{eq:tot_efield})
is a random, horizontal electric field $\bf{E}_{h}^{\rm{random}}$
given by \citep[see e.g.][]{Fan:2022}:
\begin{equation}
c {\bf{E}}_{h}^{\rm{random}} = - \nabla_{h} ( \xi B_r ) .
\label{eq:random_efield}
\end{equation}
Again, this ${\bf{E}}_{h}^{\rm{random}}$ is given as the gradient of a
scalar field and thus does not alter the observed $B_r$ on
the lower boundary (reproduced by imposing ${\bf{E}}_{h}^P$).
The formulation of this electric field is inspired by that of the
``STatistical InjecTion of Condensed Helicity" (``STITCH'')
electric field \citep{Mackay:etal:2014, Dahlin:etal:2022}
used for modeling helicity condensation and filament channel formation.
But here, $\xi$ used in equation (\ref{eq:random_efield})
is a time dependent field that is made up of a superposition
of 15721 randomly placed cells of opposite sign values. Each cell has a
2D Gaussian profile with a size scale of 1.74 Mm and a peak
amplitude of $4.07 \times 10^{13} \rm{cm^2/s}$, and varies temporally
as a sinusoidal function with a period and life time of 11.89 min.
Figure \ref{fig:twistemf}(e) shows a snapshot of the $\xi$ field, which
illustrates its random cellular pattern, and the resulting random electric
field components $E_{\theta}^{\rm random}$ (Figure \ref{fig:twistemf}(f)),
and $E_{\phi}^{\rm random}$ (Figure \ref{fig:twistemf}(g)) produced at
the lower boundary when this $\xi$ field and the $B_r$ given
in Figure \ref{fig:twistemf}(a) are applied in
equation (\ref{eq:random_efield}).
As is described in \citet{Fan:2022},
${\bf{E}}_{h}^{\rm{random}}$ effectively drives random rotations
of the foot points of the $B_r$ flux concentrations, with positive (negative)
$\xi$ corresponding to clockwise (counter-clockwise) rotation.
Here the positive and negative signed cells in the $\xi$ field are
statistically balanced, therefore no net twist or helicity is driven into
the corona by ${\bf{E}}_{h}^{\rm{random}}$.
It represents the effect of turbulent convection
that drives field-line braiding which produces resistive and viscous
heating in the corona.
Similar ways of driving coronal heating by imposing random foot-point
motions that represent turbulent convection at the photospheric lower
boundary of MHD simulations have been widely used
\citep[e.g.][]{Gudiksen:etal:2005,Warnecke:Peter:2019}.

Even though the driving electric field is derived from the
photosphere magnetograms, we impose a fixed chromosphere temperature of 20,000 K
and density of $10^{12} \, {\rm cm}^{-3}$ at the lower boundary.
The boundary conditions for the side and top boundaries in the present
simulation are the same as those used in \citet{Fan:2017}.

\subsection{The initial state}
\label{sec:init_state}
\begin{figure}[htb!]
\centering
\includegraphics[width=0.5\textwidth]{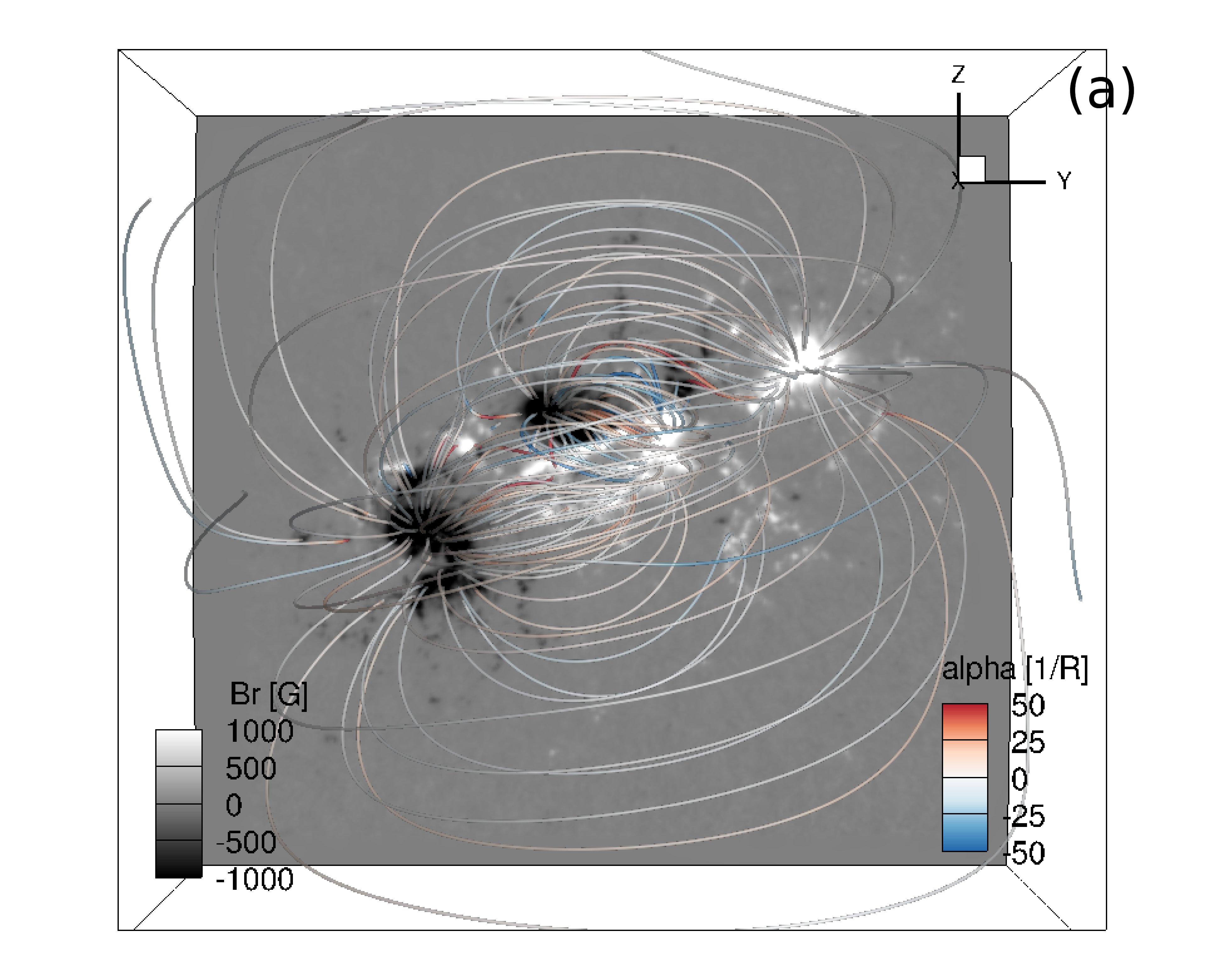} \\
\includegraphics[width=0.5\textwidth]{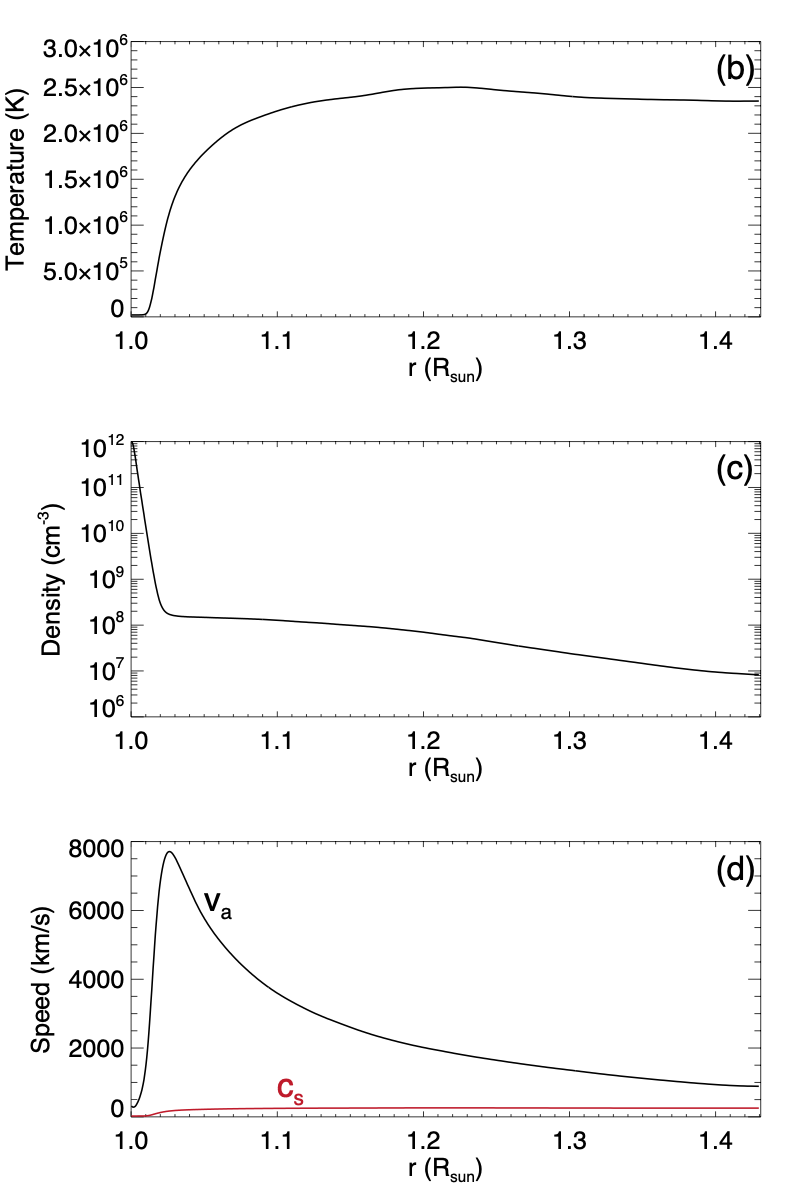}
\caption{3D field lines of the magnetic field colored with the twist
rate $\alpha$ (a) (see text for the definition of $\alpha$), and the radial
profiles of the horizontally averaged temperature (b), density (c), and
the Alfv\'en and sound speeds (d) for the initial state of the simulation.}
\label{fig:initstate}
\end{figure}
To set up the initial state of the data-driven simulation, we start with a
potential magnetic field extrapolated from the observed photospheric normal
magnetic field $B_r$ at 2011-02-14 23:47 UT (Fig. \ref{fig:twistemf}(a),
about 1.9 hours before the onset of the observed
X-class flare), and numerically evolve the MHD state by driving
at the lower boundary with only the random electric field
${\bf{E}}_{h}^{\rm{random}}$ until it reaches a
quasi-steady state with a hot corona.
Figure \ref{fig:initstate} shows the 3D magnetic field lines (panel (a))
colored with the twist rate defined as
$\alpha \equiv {\bf J} \cdot {\bf B} / B^2$ where
${\bf J} = \nabla \times {\bf B} $ is the current density, and the
radial profiles of the horizontally averaged temperature (panel (b)),
density (panel (c)), and the Alfv\'en and sound speeds (panel (d))
of the relaxed, quasi-steady state reached.
It can be seen that the magnetic field (panel (a)) for the relaxed state
is no longer the potential field but contains some random twist $\alpha$
due to the driving random electric field (which does not drive a net helicity
into the domain),
although most field lines remain near zero
$\alpha$ (white color), and the magnetic energy remains very close to
that of the potential field energy (being about 1.0016 of the potential field
energy). A quasi-steady atmosphere with 
a reasonable temperature and density stratification (panels (b) and (c))
that contains a chromosphere transitioning into a hot corona is established.
The Alfv\'en speed is significantly higher than the sound speed throughout
the domain (panel (d)).

We then use this relaxed state as the initial state (defined hereafter
as $t=0$) for the simulation, driven at the lower boundary with all
three electric fields given in the right hand side of
equation (\ref{eq:tot_efield}) (derived from a time sequence of the
vector magnetograms starting from 2011-02-14 23:47 UT) for a perid of
over 2.8 hours.
The following section describes the resulting evolution obtained from the
boundary data-driven simulation.

\section{Simulation results}
\label{sec:result}

\subsection{Overview of evolution}
\label{sec:overview_evolution}
\begin{figure}[htb!]
\centering
\includegraphics[width=0.49\textwidth]{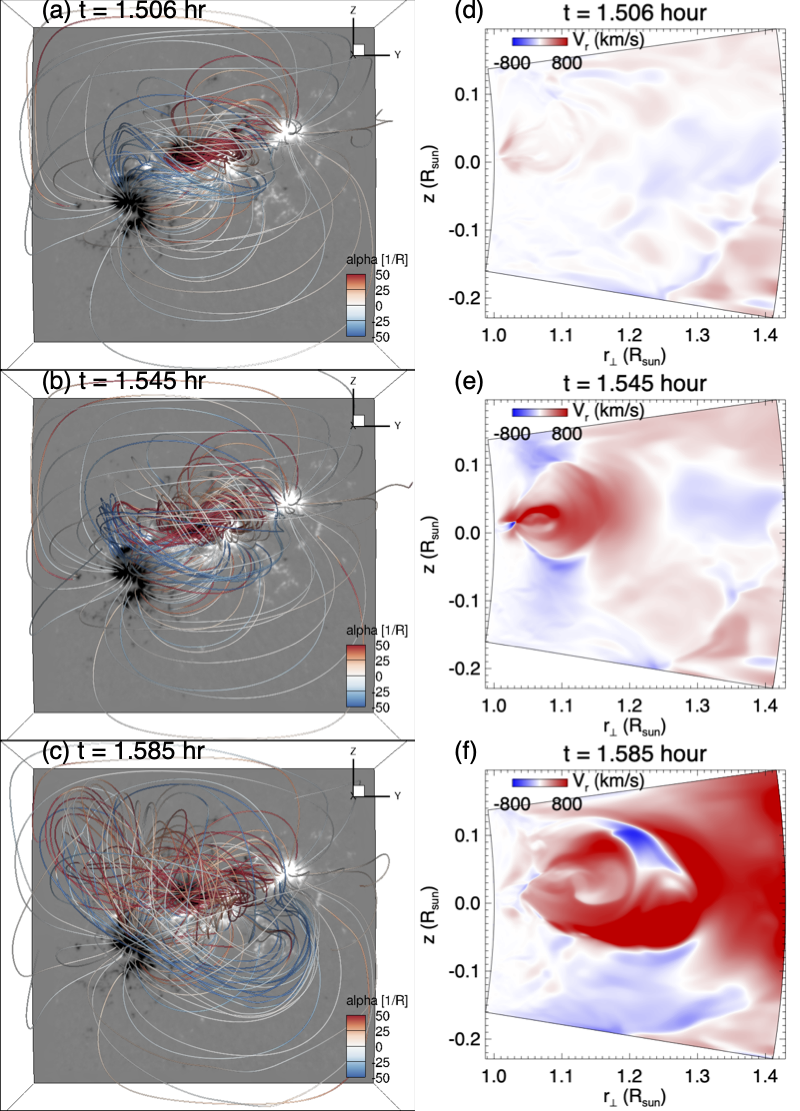}
\includegraphics[width=0.46\textwidth]{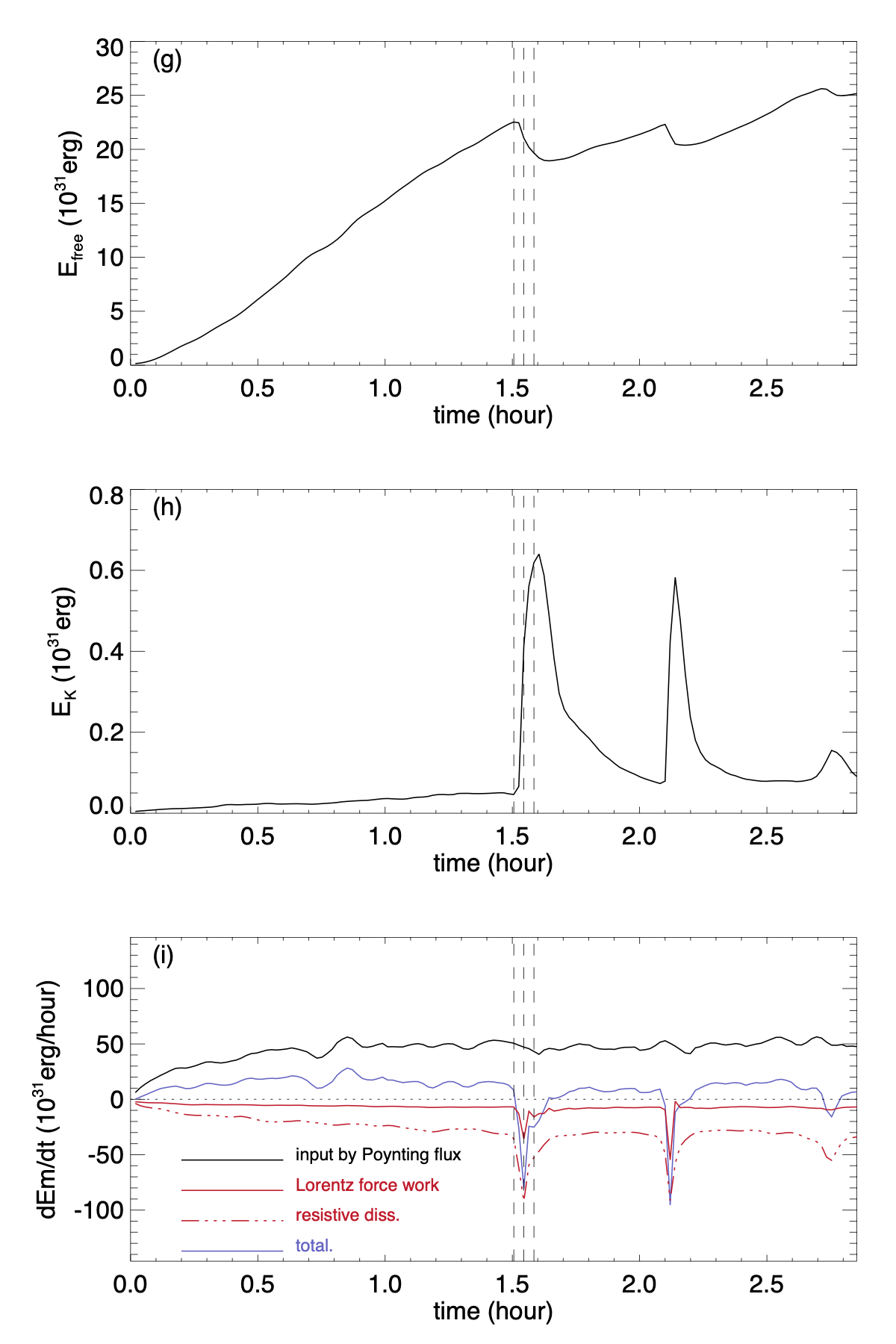}
\caption{(a,b,c) Snapshots of the magnetic field evolution showing 3D field
lines colored with the twist rate $\alpha$ at three time instances during the
onset of the first eruption,
(d,e,f) snapshots of the radial velocity in the middle meridional
cross-section of the domain at the corresponding times of the 3D magnetic
field images in the left column, and (g) the evolution of the free magnetic
energy $E_{\rm free}$, (h) the evolution of the kinetic energy $E_k$,
and (i) the contributions to the rate of change of the magnetic energy due to
the input from the integrated Poynting flux at the boundary (black curve),
the total Lorentz force work (red solid curve),
the total resistive dissipation (red dash-dotted curve),
and the sum of the three (blue curve) which corresponds to the total
rate of change of the magnetic energy. The 3 vertical dashed lines in
panels (g)(h)(i) mark the times of the snapshots displayed in the
left columns. An animated version of the figure is available online,
which shows the whole course of the 2.8-hour simulated evolution,
from the build up of the pre-eruption magnetic field through the
multiple eruptions. See text for details.}
\label{fig:emepek_demdt}
\end{figure}

Figure \ref{fig:emepek_demdt} shows the resulting evolution of the 3D
magnetic field (left column images), the radial velocity in the central
meridional cross-section (middle column), and several integrated
quantities (right column)
that include the free magnetic energy $E_{\rm free}$ (panel (g)),
which is the excess of the total magnetic energy over that of the
corresponding potential magnetic field, the total kinetic energy
$E_k$ (panel (h)), and the various
contributions to the rate of change of the total magnetic energy (panel (i))
which are described in the following.
(1) The black curve in panel (i) shows the input of magnetic energy by the
Poynting flux integrated over the lower and upper boundaries:
$\int (c/4\pi) ({\bf E}_h \times {\bf B}_h)_r \, dS$,
where $(c/4\pi) ({\bf E}_h \times {\bf B}_h)_r$ is the radial component of the
Poynting flux density and the integration $dS$ is over the area of the lower and
upper boundaries (note no Poynting flux through the side boundaries due to the conducting wall
side boundary condition).
(2) The solid red curve shows the release of magnetic energy resulting
from the Lorentz force work:
$- \int (1/4 \pi) ((\nabla \times {\bf B}) \times {\bf B}) \cdot {\bf v} \, dV$,
where $(1/4 \pi) ((\nabla \times {\bf B}) \times {\bf B})$ is the 
Lorentz force, ${\bf v}$ is the velocity, and the integration $dV$ is over the
volume of the domain.
(3) The red dash-dotted curve shows the dissipation of magnetic energy
by the numerical magnetic diffusion:
$- \int (c/4\pi) {\bf E}_{\rm num} \cdot (\nabla \times {\bf B}) \, dV$,
where ${\bf E}_{\rm num}$ is the electric field resulting from the
numerical diffusion evaluated in the numerical code.
The blue curve shows the sum of the above three, which is the net
rate of change of the total magnetic energy.
We see an overall continuous build up of the free magnetic energy $E_{\rm free}$
(panel (g)) due to the continuous Poynting flux input (black curve in panel (i))
at the lower boundary produced by the driving electric field (${\bf E}_h$ given by
eq. [\ref{eq:tot_efield}]).
This Poynting flux input is in excess of the dissipations of the magnetic energy due to
the resistive dissipation (red dash-dotted curve in panel (i))
and the Lorentz force work (red solid curve in panel (i) ),
resulting in a continuous net gain
of the free magnetic energy (blue curve in panel (i))
for most of the time.
But this continuous build up is punctuated by sudden free magnetic energy
$E_{\rm free}$ releases and kinetic energy $E_k$ increases (at about
$t=1.51$ hour, $t=2.1$ hour, and $t=2.7$ hour), due to the sudden
enhancements of the Lorentz force work and resistive dissipation resulting from
the loss of equilibrium of the magnetic field (see the movie for the whole
course of the evolution).

During the first 1.5 hour period, $E_{\rm free}$ increases
steadily, while $E_k$ and the Lorentz force work remain small with the
magnetic field being in quasi-equilibrium.
We can see from the movie of Figure \ref{fig:emepek_demdt}
that during this time the magnetic field is being sheared and
twisted as indicated by the color of the field lines, forming
positive-twisted (red field lines) sigmoidal shaped loops above the
central (PIL) and negative-twisted (blue field lines) inverse-S shaped
loops connecting to the following negative polarity sunspot (panel (a)).
At about $t=1.506$ hour,
the free magnetic energy reaches a peak value of about $2.25 \times
10^{32} {\rm erg}$ (panel (g)),
and then we see the onset of the first eruption with a
sudden outward acceleration of the central sigmoid field
(panels (b) and (e)), forming a positive-twisted
erupting flux rope (red twisted field lines) which also pushes
out and accelerates an outer negative-twisted field (blue field
lines) (panels (c) and (f) and the movie).

\subsection{The Pre-eruption Magnetic Field}
\label{sec:preeruption}
\begin{figure}[htb!]
\centering
\includegraphics[width=0.5\textwidth]{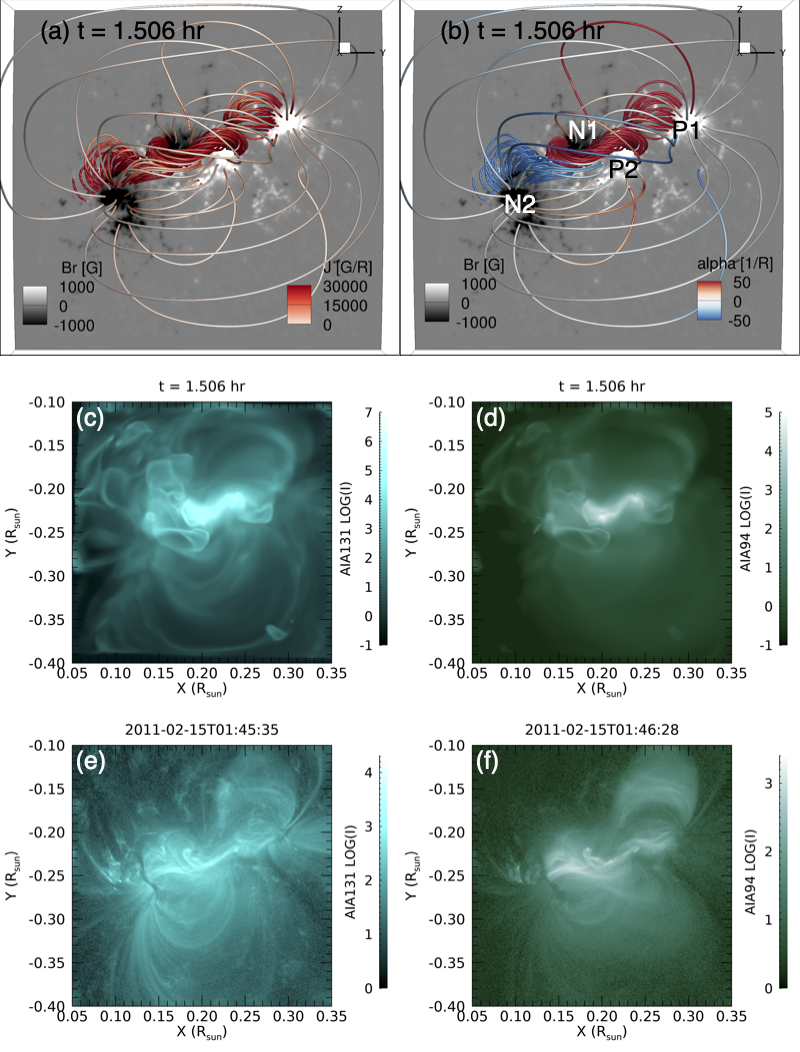}
\caption{Top 2 panels show selected 3D magnetic field lines of the
pre-eruption magnetic field just before the onset of the eruption at
$t=1.51$ hour, with the field lines colored in current density $J$ (a)
and in twist rate $\alpha$ (b). The middle
row panels show the synthetic AIA 131 {\AA} image (c) and synthetic
AIA 94 {\AA} image (d) at the same time.
The bottom panels show in comparison the
observed AIA 131 {\AA} image (e) and 94 {\AA} image (f)
at respectively 2011-2-15 01:45:35 UT and 2011-02-15 01:46:28,
which are about the time for the onset of the observed X2.2 flare.}
\label{fig:preeruption}
\end{figure}

Figure \ref{fig:preeruption} shows the pre-eruption magnetic field
just before the onset of the eruption at $t=1.506$ hour. The top two
panels show a set of 3D magnetic field lines colored with the
current density $J$ (panel (a)) and the twist rate $\alpha$ (panel (b)).
Comparing the pre-eruption field in Figure \ref{fig:preeruption}(b)
 to the nearly potential field of the initial state in
Figure \ref{fig:initstate}(a), we find that the twisting electric field
has built up highly sheared, forward S-shaped (sigmoid) loops with positive
(right-handed) twist rates $\alpha$ above the central PIL
connecting polarities P2, N1,
and also positive-twisted loops connecting polarities P1, N1.
Furthermore, we find negative-twisted (left-hand-twisted) loops
(blue with negative $\alpha$) connecting P2, N2 polarities, and
some higher loops with negative twist connecting P1, N2
polarities. As pointed out earlier in Figure \ref{fig:twistemf},
the twisting electric field imposed in N2 (with opposite signs
of $J_r$ and $B_r$) transports negative twist into the corona,
opposite to the twisting electric field dominating in the other
polarity concentrations that drives positive twist into the
corona.

It can be seen in Figure \ref{fig:preeruption}(a) that the
field lines with the strongest current density $J$
show morphology in qualitative agreement with the brightening loops
in the observed hot channel images of SDO/AIA
(Figs. \ref{fig:preeruption}(e) and \ref{fig:preeruption}(f)).
This indicates that the model captures the structure
of the non-potential (energized) magnetic field.
Figures \ref{fig:preeruption}(c) and \ref{fig:preeruption}(d)
show respectively the synthetic AIA 131 {\AA} and 94 {\AA} images
at the same time of the pre-eruption magnetic field shown
in Figures \ref{fig:preeruption}(a) and \ref{fig:preeruption}(b).
The synthetic images are computed by integrating along the line of
sight (from the Earth perspective) through the simulation domain:
\begin{equation}
I = \int n_e^2 (l) \, f(T(l)) \, dl,
\end{equation}
where $l$ denotes the length along the line of sight through the
simulation domain, $I$ denotes the intensity at each pixel in
units of DN/s/pixel (shown in LOG scale in the images),
$n_e$ is the electron number density, and $f(T)$ is the temperature
response function of the corresponding AIA filter (obtained by
running the SolarSoft routine {\tt get\_aia\_response.pro}).
The synthetic AIA images
(Figs. \ref{fig:preeruption}(c) and \ref{fig:preeruption}(d))
also show some qualitatively similar emission features as those
in the observed ones
(Figs. \ref{fig:preeruption}(e) and \ref{fig:preeruption}(f)),
i.e. the central forward S-shaped sigmoid emission along
the central PIL, and the surrounding large loops connecting the
major polarity concentrations.
We find that the central sigmoid emission is due to the
heating produced by the formation of a strong current layer
in the strongly sheared field above the central PIL.

\begin{figure}[htb!]
\centering
\includegraphics[width=0.3\textwidth]{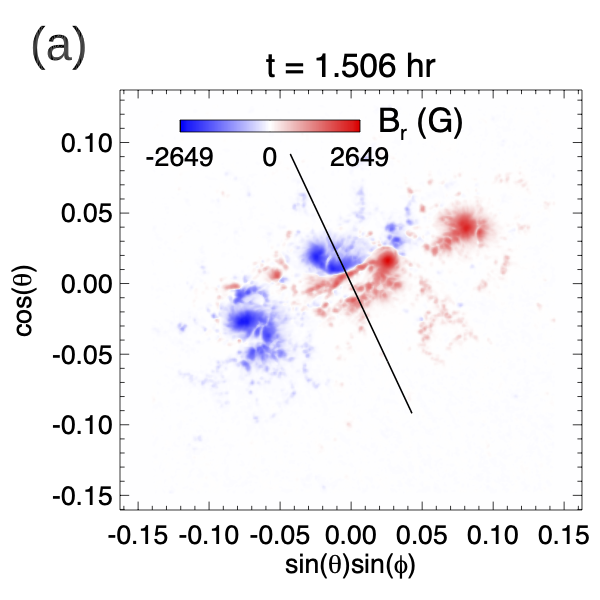}
\includegraphics[width=0.3\textwidth]{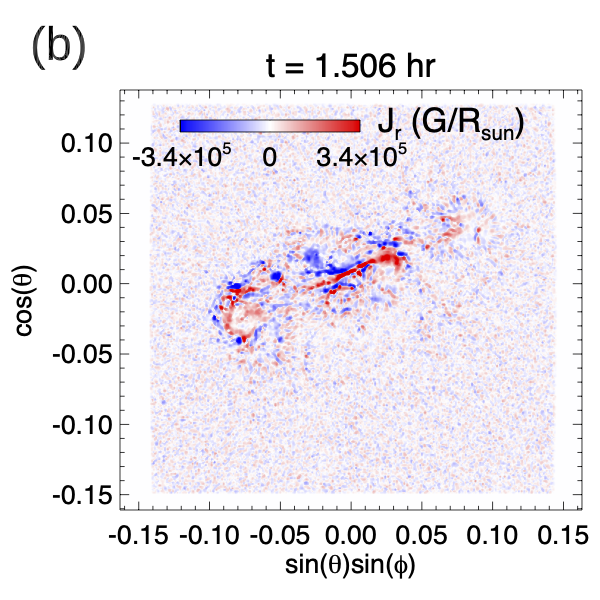}
\includegraphics[width=0.3\textwidth]{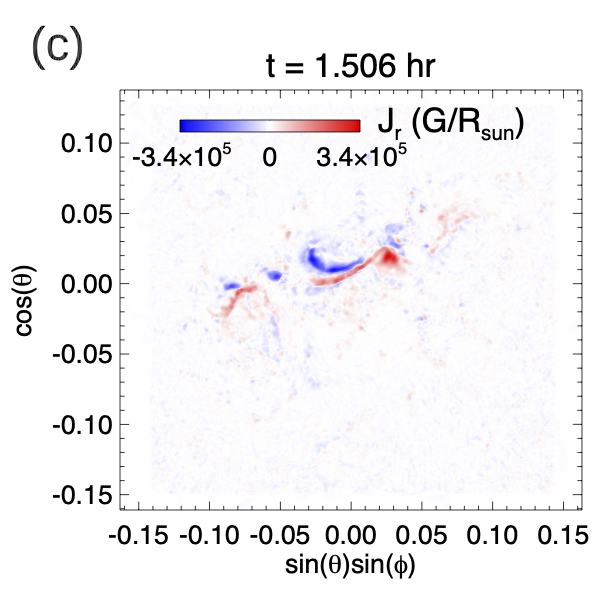}
\caption{(a) The observed vertical magnetic field $B_r$ and (b)
the vertical electric current density $J_r$ measured from the vector
magnetogram, compared with (c) the vertical electric current density
$J_r$ obtained at the lower boundary of the simulation domain at
time $t = 1.506$ hour, about the time of the onset
of the first eruption in the simulation.}
\label{fig:comparejr}
\end{figure}

To examine how close to force-free the pre-eruption magnetic
field is, we have evaluated $\sigma_J$, the current-weighted
mean of the sine of the angle $\theta$ between the current
density ${\bf J}$ and the magnetic field ${\bf B}$
\citep{Wheatland:etal:2000}:
\begin{equation}
   {\sigma}_J = \frac{\sum J_i \sin \theta_i}{\sum J_i}
\end{equation}
where
\begin{equation}
   \sin \theta_i = \frac{|{\bf J} \times {\bf B}|_i}{J_i B_i},
\end{equation}
the subscript ``i'' denotes each grid point
and the sum is over all the grid points in the simulation
domain.  We found $\sigma_J \approx 0.06 \ll 1$, corresponding to
a (current-weighted) mean angle of about $3.4^{\circ}$ between the
current density and the magnetic field vectors, i.e. close to
being force-free. Furthermore we found that the vertical current
density obtained at the simulation lower boundary is close to the
observed $J_r$ from the HMI vector magnetogram at the corresponding
time, for the large-scale main current patches,
as can be seen in Figure \ref{fig:comparejr} (compare panels (b)
and (c)).
This shows that the modeled pre-eruption magnetic field is close to
the solution of a NLFFF extrapolation
from the observed vector magnetogram.
Indeed, the peak free energy of the pre-eruption magnetic field
reached just before the first eruption ($2.25 \times
10^{32} {\rm erg}$) is close to the result obtained in
\citet{Sun:etal:2012}, who carried out the NLFFF
extrapolation of the coronal magnetic field of AR 11158 and
found that the free magnetic energy reaches a maximum of
$\sim 2.6 \times 10^{32} {\rm erg}$ just befored the onset
of the observed X2.2 flare.
As described in \S \ref{sec:efield}, the twisting electric field
we impose does not enforce the observed horizontal magnetic
field at the lower boundary.
Nevertheless, we find that the resulting horizontal magnetic
field at the lower boundary for the pre-eruption magnetic field
that is built up is very similar to that of the observed.
\begin{figure}[htb!]
\centering
\includegraphics[width=0.9\textwidth]{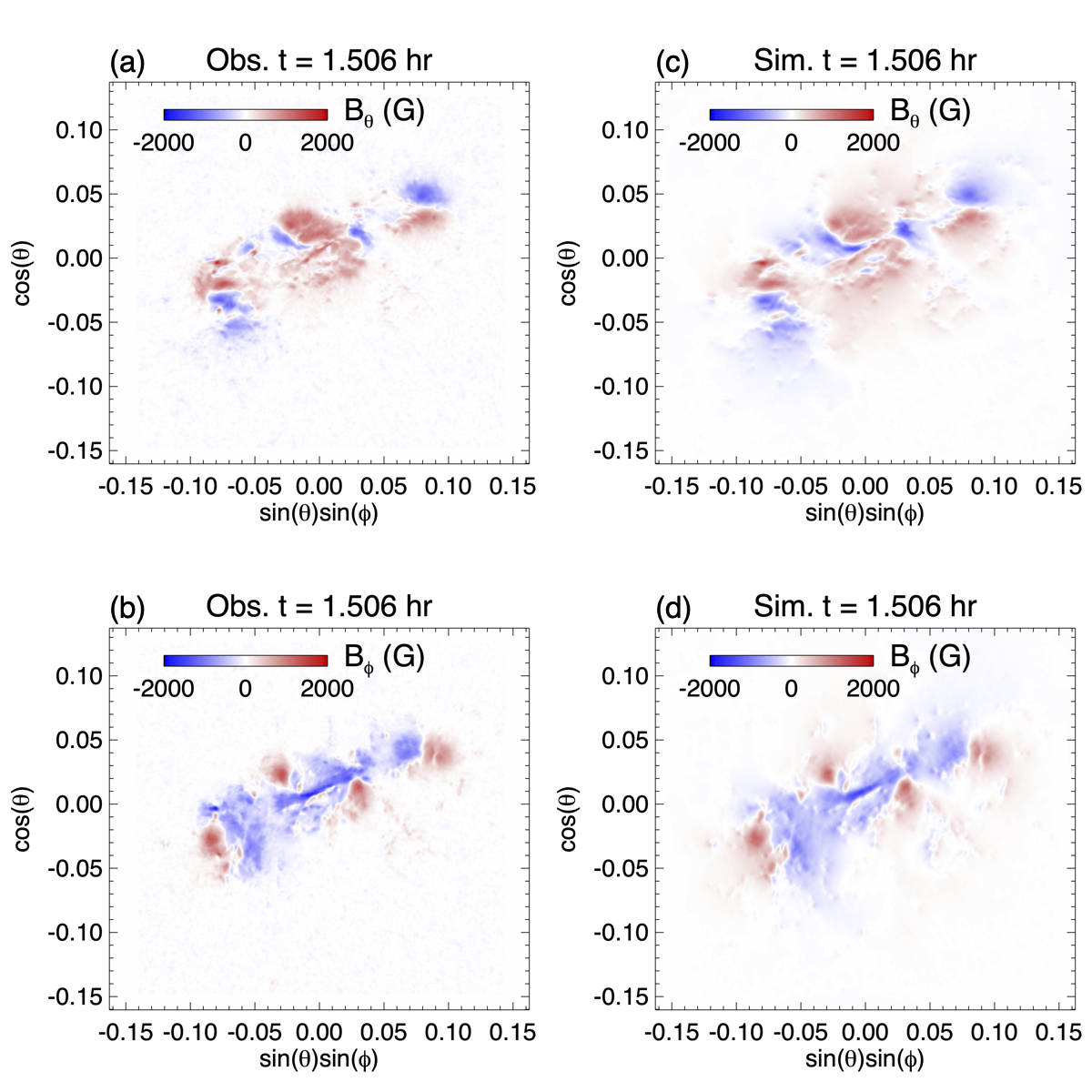}
\caption{The observed photospheric horizontal components of the magnetic
field $B_{\theta}$ (a) and $B_{\phi}$ (b) compared with the lower
boundary $B_{\theta}$ (c) and $B_{\phi}$ (d) resulting from the simulation
at the time of the onset of the simulated eruption.}
\label{fig:compare_Bh}
\end{figure}
Figure \ref{fig:compare_Bh} shows a comparison of the
horizontal components of the magnetic field at the lower boundary
at the onset of the simulated eruption with those from the
photospheric vector magnetogram at the same time. We see very
similar patterns for both the $B_{\theta}$ and $B_{\phi}$ components
comared to the observed ones.

\subsection{The initiation of eruption}
\label{sec:initiation}
\begin{figure}[htb!]
\centering
\includegraphics[width=0.85\textwidth]{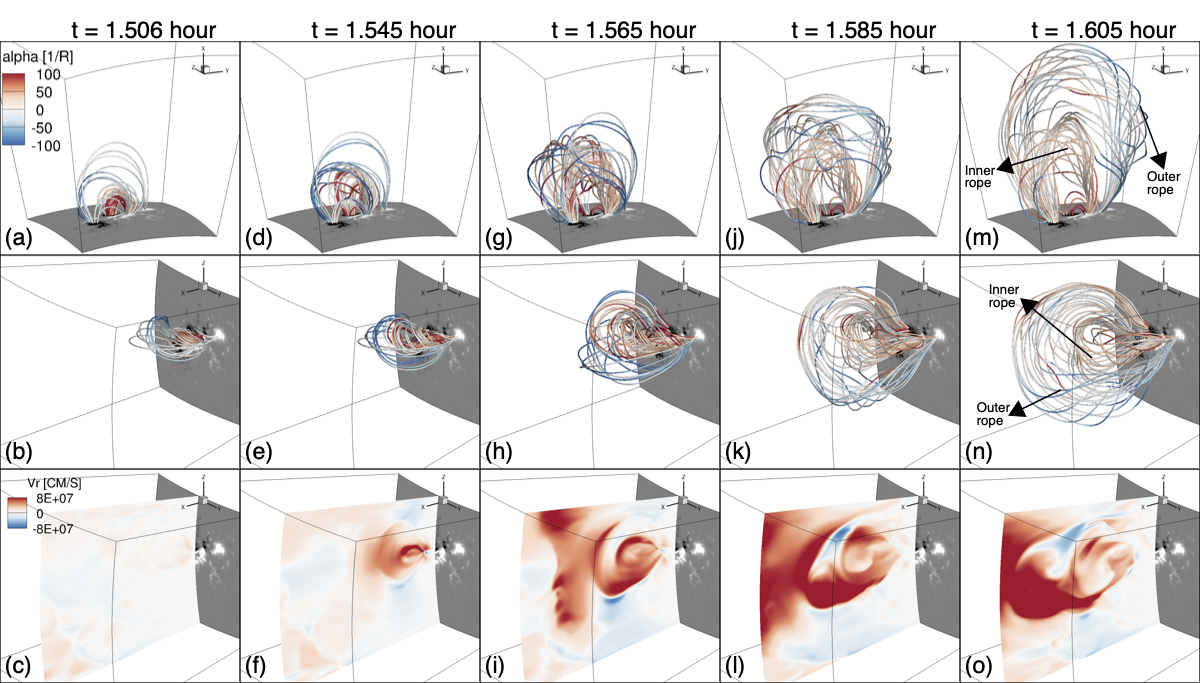}
\caption{The top 2 rows show snapshots of the 3D evolution of the erupting magnetic
field viewed from 2 different perspectives during
the first eruption. The set of field lines are traced from a set of Lagrangian
tracer points tracked in the velocity field.
The field lines are colored in the twist rate $\alpha$.
The bottom row shows the evolution of the radial velocity
in the central meridional cross-section, with the same perspective view
as the 2nd row.
The different columns correspond to different time instances with time
running from left to right. An animated version of this figure is
available online,
which shows the 3D magnetic field evolution during the initial 7 min of the
first eruption.}
\label{fig:evolerupt}
\end{figure}
\begin{figure}[htb!]
\centering
\includegraphics[width=0.45\textwidth]{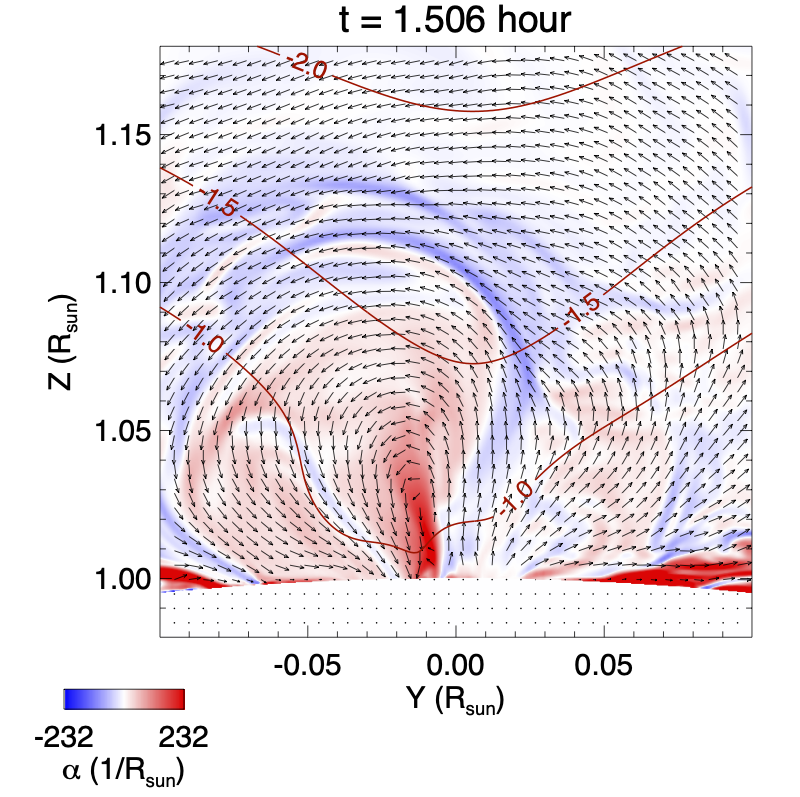}
\includegraphics[width=0.45\textwidth]{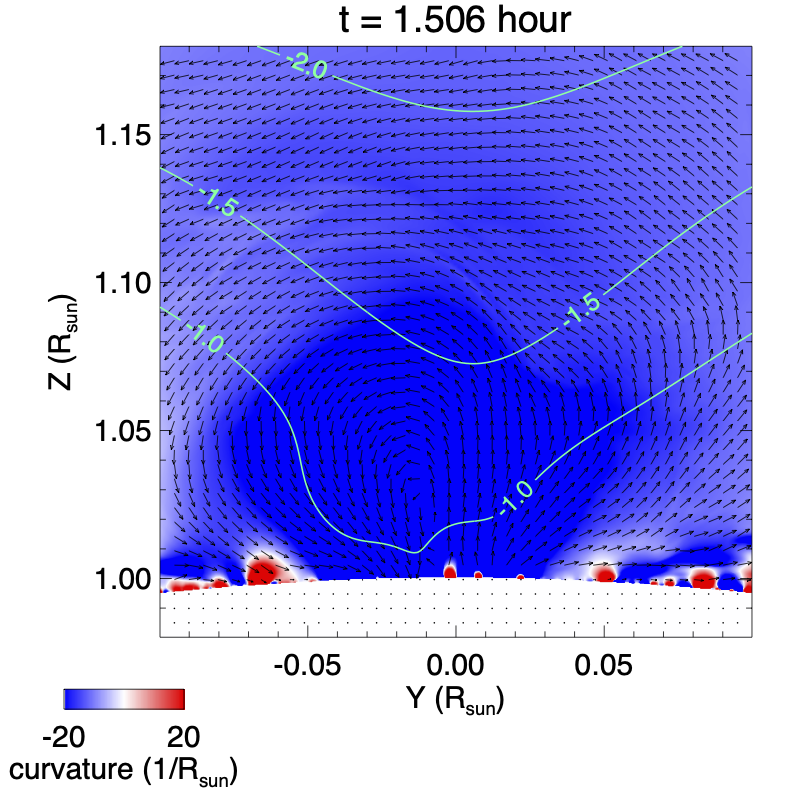} \\
\includegraphics[width=0.45\textwidth]{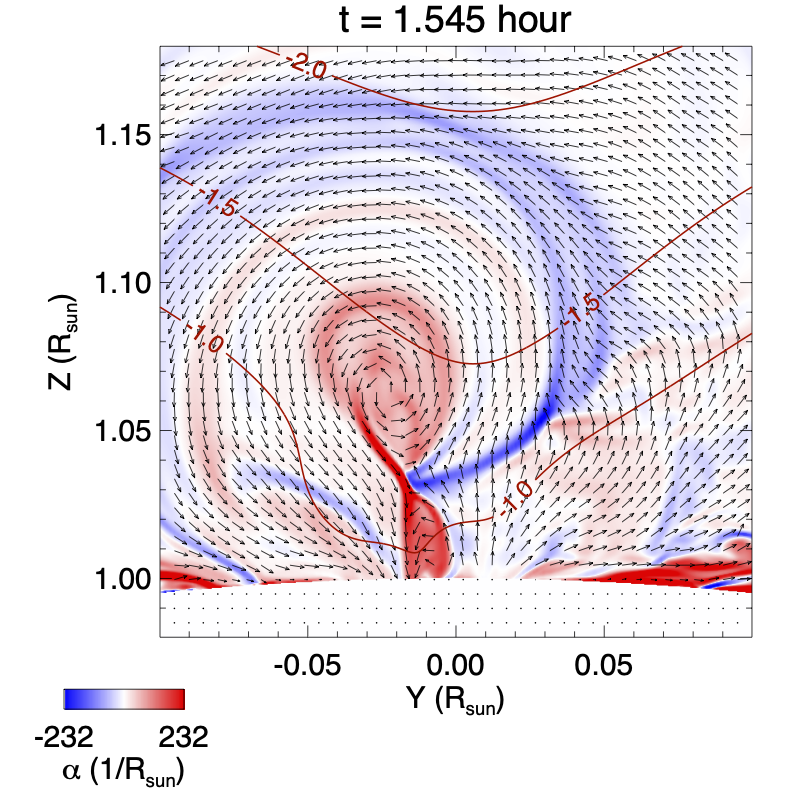}
\includegraphics[width=0.45\textwidth]{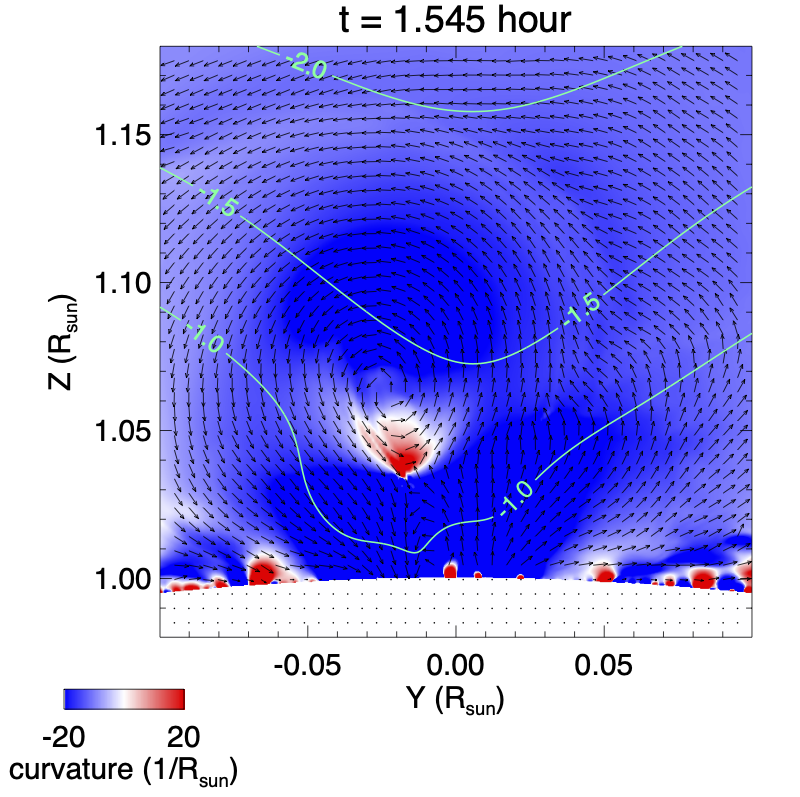}
\caption{Twist rate $\alpha$ (left panels) and the
radial curvature ${\bf B} \cdot \nabla (B_r) / B^2$ (right panels) in the vertical
cross-section (whose location is indicated by the black line in
Figure \ref{fig:comparejr}(a)) across the middle of the central PIL
at the time of the onset of the first eruption (top) and at a time
during the eruption (bottom), overlaid with the arrows that show the direction
of the magnetic field within the cross-section plane, and also overlaid with the
contours of the decay index $d ({\rm ln} B_{p{\rm h}})/d ({\rm ln} h)$,
where $B_{p{\rm h}}$ is the horizontal field strength of the corresponding
potential field and $h$ denotes the height above the lower boundary surface.}
\label{fig:alpha_curvature_bpdecay}
\end{figure}

Figure \ref{fig:evolerupt} shows the 3D evolution of a set of
erupting magnetic field lines as viewed from 2 different
perspectives (top 2 rows) during the first eruption.
The field lines are traced from a set of Lagrangian tracer points tracked
in the velocity field. The eruption begins with the acceleration of the
highly sheared sigmoid field with positive twist
above the central PIL (see the red field lines in panel (a)
and the sudden onset of the rise velocity from panel (c) to panel (f)
of Fig. \ref{fig:evolerupt}).
To examine what might have triggered the sudden acceleration of the sigmoid
field above the central PIL and whether it contains a magnetic flux rope,
we plotted in Figure \ref{fig:alpha_curvature_bpdecay} the distribution
of twist rate $\alpha$ and the radial curvature
${\bf B} \cdot \nabla (B_r) / B^2$ in the vertical cross-section, whose
location is indicated by the black line across the central PIL
in Figure \ref{fig:comparejr}(a), at the time of
the onset of the eruption ($t=1.506 $ hour, top row) and at a time
during the eruption($t=1.545$ hour, bottom row).
It can be seen in the top left panel that most of the current of
the positive-twisted (red) sigmoid field region is still below the contour
of the critical decay index of $d ({\rm ln} B_{p{\rm h}}) / d ({\rm ln} h) = -1.5$
for the onset of the torus instability \citep[e.g.][]{Kliem:Toeroek:2006}.
Here $B_{p{\rm h}}$ is the horizontal field strength of the corresponding
potential field and $h$ denotes the height above the surface.
In the mean time, it can also be seen that a thin
current layer (thin layer of strong positive $\alpha$) has developed
just above the PIL (located at about $Y=-0.01$ in
Figure \ref{fig:alpha_curvature_bpdecay})
in the sigmoid field region, which indicates the
onset of the tether-cutting reconnection at the current layer may
be the trigger of the onset of the eruption instead of the torus instability.
Furthermore, we find no dipped field with positive radial curvature above the
central PIL (at $Y=-0.01$) in the top right panel of
Figure \ref{fig:alpha_curvature_bpdecay},
indicating that at the onset of the eruption, the sigmoid field above the PIL
is a sheared arcade instead of a flux rope.
The bottom panels of Figure \ref{fig:alpha_curvature_bpdecay} corresponding
to shortly after eruption onset show that
a flux rope with dipped field lines with positive radial curvature 
forms during the eruption (bottom right panel) as a result of the
tether-cutting reconnection in the vertical current sheet above the
central PIL (bottom left panel).
\begin{figure}[htb!]
\centering
\includegraphics[width=0.8\textwidth]{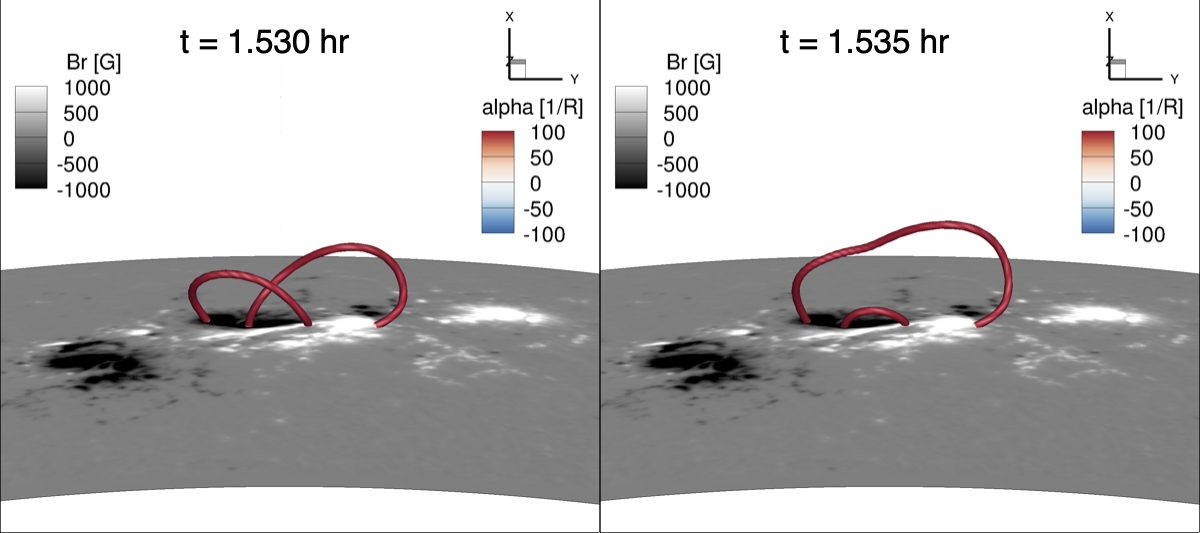}
\caption{3D view of example field lines before (left panel) and
after (right panel) undergoing tether-cutting reconnections. Field lines are
colored in the twist rate $\alpha$ and the gray scale image shows the normal
magnetic field $B_r$ in the lower boundary.}
\label{fig:3Dreconnect}
\end{figure}
The onset of the tether-cutting reconnection is facilitated by the
numerical resistivity, triggered by the thinning
of the current sheet in the strongly sheared arcade field above the PIL.
Figure \ref{fig:3Dreconnect} shows a 3D view of example
field lines before and after undergoing tether-cuttig reconnections.  
We see that a pair of highly sheared arcade field lines above the central PIL
in the earlier time instance (left panel) have reconnected and transformed
into a longer dipped field line that erupts upward and a lower short loop that 
shrinks back down in the later time instance (right panel).
The formation of a flux rope with
dipped field lines can also be seen in panels (d) and (g) in
Figure \ref{fig:evolerupt}.
As the positive-twisted inner field erupts, it encounters and reconnects
with the negative-twisted field above (see top row of Fig. \ref{fig:evolerupt}).
Eventually, the erupting field develops a complex structure that consists of an
outer flux rope containing mainly negative twist (left-handed twist) and
an inner flux rope containing mainly positive twist (right-handed twist)
(see panels (m) and (n) in Fig \ref{fig:evolerupt}), with the outer
flux rope accelerating to a speed of about $800$ km/s as it approaches the top
of the domain (panels (l) and (o) in Fig. \ref{fig:evolerupt}).
It can also be seen in panels (l) and (o) in Fig. \ref{fig:evolerupt}
(and the associated movie)
that the rise velocity of the inner flux rope appears to
decelerate towards the end.  This is due to its collision and
reconnection with the overlying negative twisted flux, which forms
the outer flux rope that continues to accelerate and exits the
domain while the inner flux rope slows down as it loses twist
via reconnection with part of the outer magnetic flux.
The top boundary is an open boundary allowing plasma to flow
through. However, the side boundaries are closed wall boundaries where
both the magnetic and velocity fields become parallel to the walls. Thus
the lateral expansion and deflection of the erupting flux ropes are
artificially constrained by the side boundaries (see result in the next section).

\subsection{The observational signatures of the erupting field}
\label{sec:erupting_fields}
\begin{figure}[htb!]
\centering
\includegraphics[width=0.8\textwidth]{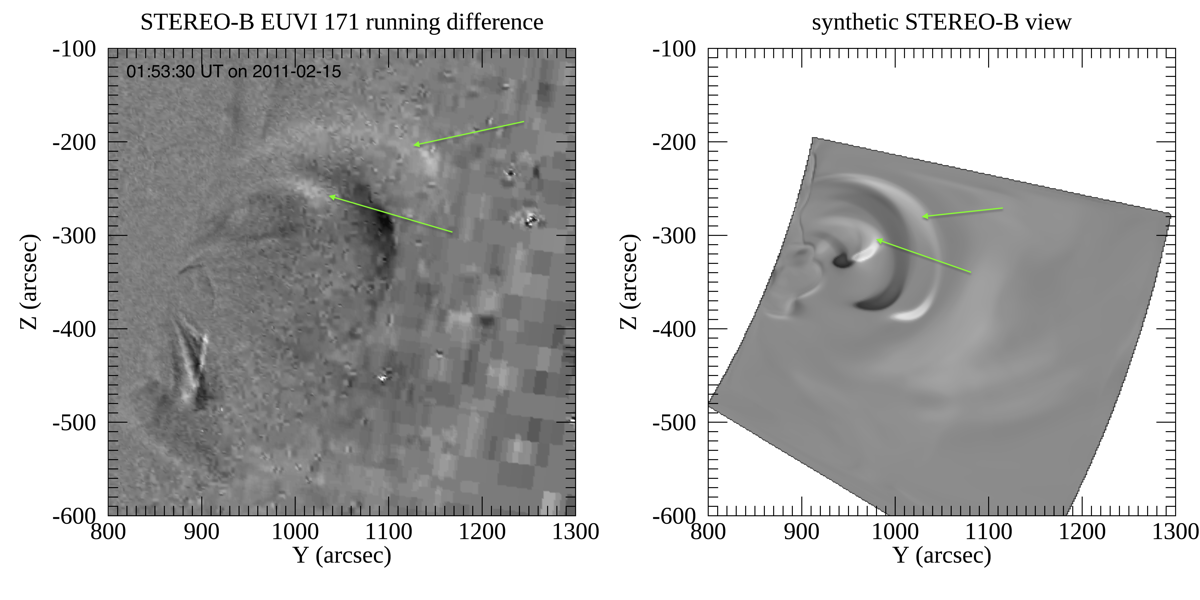}
\caption{(left) A running difference image from the STEREO-B EUVI 171 passband images
of times 01:52:15 UT and 01:53:30 UT on 2011-02-15 during the observed CME eruption,
and (right) the running difference of two synthetic images in the AIA 171 passband
but as viewed from the STEREO-B perspective at times $t=1.555$ hour
and $t=1.565$ hour during the simulated eruption.}
\label{fig:stereoB-rdiff} 
\end{figure}

To study the observational signatures of the simulated erupting field, we have
computed the synthetic EUV images in the AIA 171 {\AA} channel but viewed from
the STEREO-B perspective. The right panel of Figure \ref{fig:stereoB-rdiff} shows
the running difference of two synthetic images at times $t=1.555$ hour and $t=1.565$
hour during the eruption. It shows an outgoing double shell structure produced by
the two outgoing flux ropes. A running difference image from the observed
STEREO-B EUVI 171 passband images during the eruption (left panel of
Fig. \ref{fig:stereoB-rdiff}) shows a similar outgoing double-shell structure,
although the observed out-going structure expands side-way
significantly more than the simulated one because of the constraining side wall
boundary of the simulation domain.  Simulation with a widened domain is needed for a
quantitative comparison of the kinematics of the erupting structure.

\begin{figure}[htb!]
\centering
\includegraphics[width=0.9\textwidth]{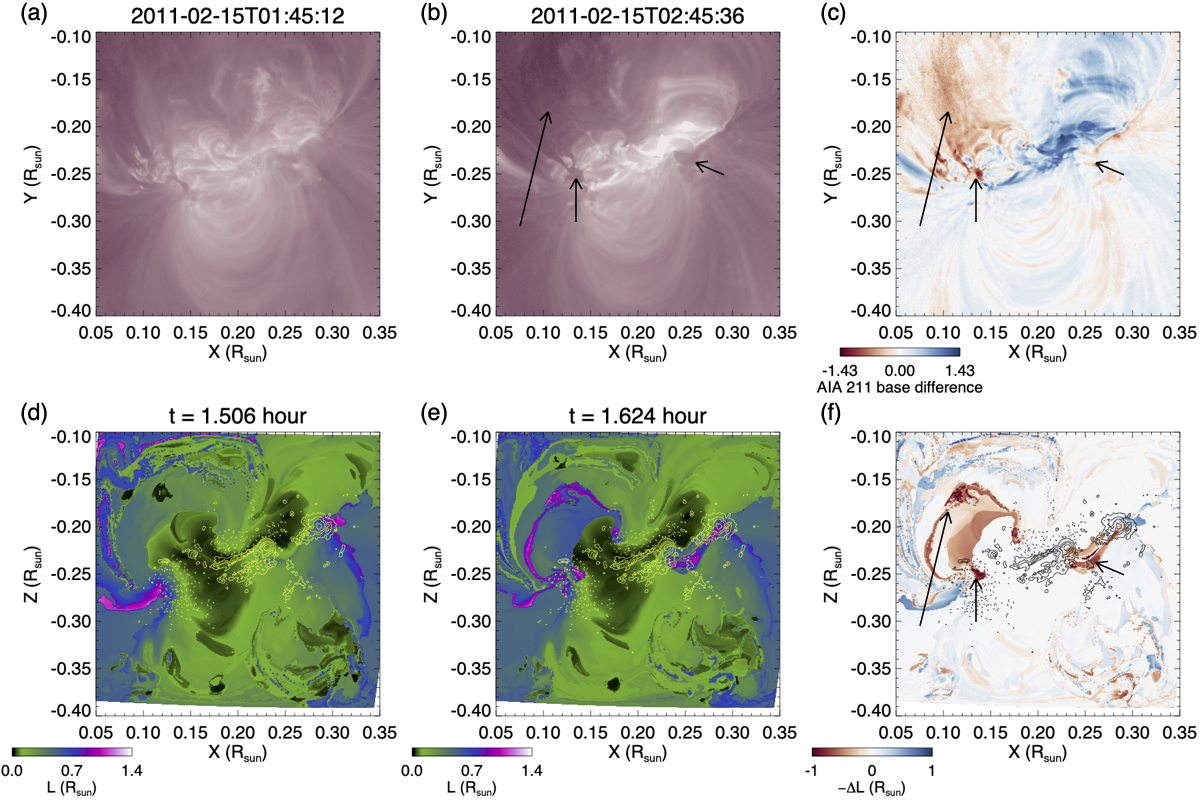}
\caption{Upper panels show show AIA 211 {\AA} channel images at the onset of (a) and
during (b) the eruption, and the difference image (c) between the two.
Lower panels show the lengths of the field lines traced from each location
at the base of the corona at the onset of the eruption (d) and at a later time when
the outer erupting flux rope is exiting the domain (e), and the change of field
line lengths between the two times (f). Three black arrows mark the locations
of the major EUV dimming regions observed. The regions marked by the two short
arrows correspond to the core dimmings and the region marked by the long arrow
correspond to a more diffuse secondary dimming \citep[e.g.][]{Dissauer:etal:2018}.
Contour lines in the bottom panels show the radial magnetic field on the photosphere,
with solid (dotted) contours corresponding to positive (negative) radial fields.}
\label{fig:dimmings}
\end{figure}

CME-associated coronal dimmings are thought to correspond to footpoints of the
expanding CME magnetic structures and caused by a density decrease in these structures
due to their rapid expansion \citep[e.g.][]{Thompson:etal:1998, Dissauer:etal:2018}.
Here we use our data-driven simulation to identify the footpoints of the erupting field
lines and compare them with
the observed EUV dimmings of the CME event. The bottom panels in
Figure \ref{fig:dimmings} show the lengths of the field lines traced from each
locations at the base of the corona at the onset of the eruption (panel (d)),
and at a time when the outer erupting flux rope is exiting the domain (panel (e)),
and the difference between the two times (panel (f)) which shows the change of the
field line lengths due to the eruption.
The regions of large field line length increases (as represented by the red patches in
panel (f)) correspond to the foot points of the stretched-out erupting field lines.
We find that these regions of the erupting field line foot points correspond well with
both the two core dimming regions (indicated by the two short arrows) as well as the
the more diffuse secondary dimming region (indicated by the long arrow) seen
in the AIA 211 {\AA} channel observations (panels (b) and (c)).
This agreement adds validation to the modeled erupting field.

\begin{figure}[htb!]
\centering
\includegraphics[width=0.45\textwidth]{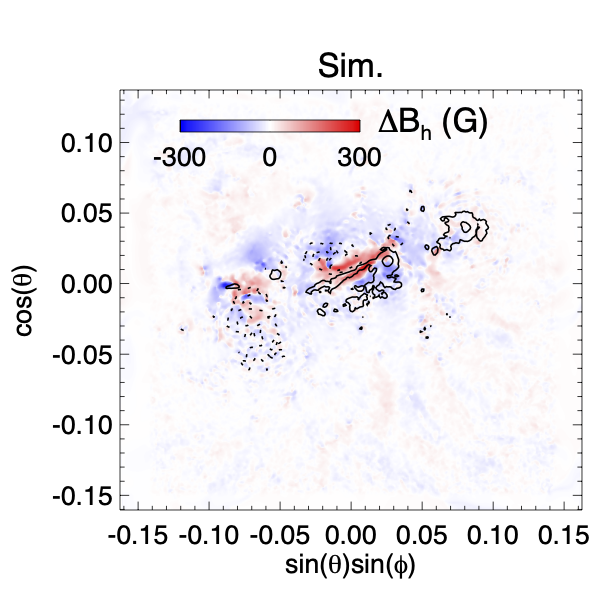}
\includegraphics[width=0.45\textwidth]{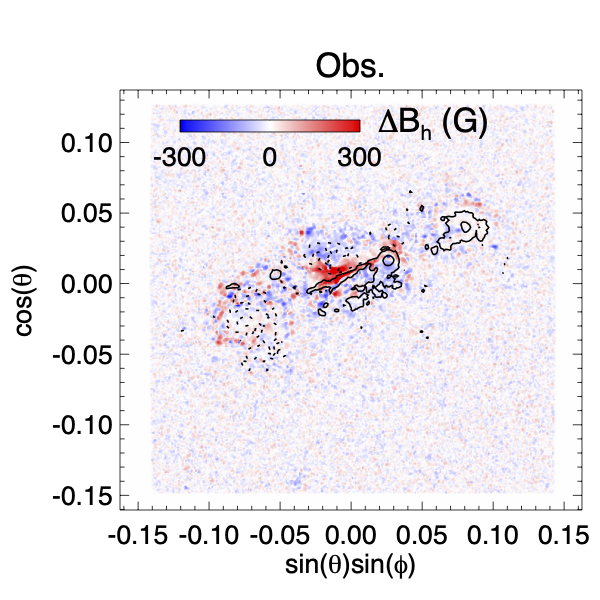}
\caption{(left) The change of the horizontal magnetic field strength $\Delta B_h$
at the lower boundary 12 min after the onset of the first eruption in the simulation,
and (right) the observed change $\Delta B_h$ in the photosphere vector magnetograms
12 min after the onset of the observed X-class flare. The black lines are contours of the
normal magnetic field $B_r$ with solid (dotted) lines corresponding to
positive (negative) $B_r$.}
\label{fig:bhchange}
\end{figure}

Observations have also shown that there is a rapid and irreversible change of the
photospheric magnetic field associated with a solar flare \citep[e.g][]{Wang:etal:1994,
Sun:etal:2012,Liu:etal:2012} as a result of the
``implosion" of the coronal magnetic field just above the flare PIL
due to the release of the magnetic energy \citep{Hudson:2000,Fisher:etal:2000a}.
In Figure \ref{fig:bhchange} we show the change of the horizontal magnetic field
strength $\Delta B_h$ at the lower boundary 12 min after the onset of the eruption
in the simulation compared to the observed change of the horizontal magnetic field
strength in the photosphere vector magnetogram 12 min after the onset of the
observed X-class flare.
We find that the change of $B_h$ in the simulation shows a similar pattern as the
observed change in the central flaring region, where there is a significant
enhancement of $B_h$ over the central PIL due to the implosion of the
post reconnection loops, while there is a systematic decrease
of $B_h$ in the periphery of the flare region away from the central PIL, where
the field becomes more vertical as it erupts.
This agreement is further evidence that our simulated eruption qualitatively
reproduces the magnetic field reconfiguration during the flare.

\section{Summary and Discussions}
\label{sec:summary}
We have performed a boundary data-driven MHD simulation of 
the eruptive flare and CME that occurred on 2011-02-15 from AR 11158.
The simulation is driven at the lower boundary with an electric
field derived from the observed evolution of the normal magnetic
field $B_r$ and the vertical electric current density $J_r$ measured
from the HMI vector magnetograms as described in \S \ref{sec:efield}.
In the simulation, the twisting electric field based on the observed vertical
electric current energizes the initial potential field to build up a
pre-eruption coronal magnetic field that is close to the NLFFF
extrapolation and it subsequently develops multiple eruptions.
The sheared/twisted field
lines of the pre-eruption magnetic field show morphologies in
qualitative agreement with the brightening loops observed in the
SDO/AIA hot passband images.
From the simulation we find that the eruption is initiated by
the tether-cutting reconnection of the highly sheared sigmoid
field above the central PIL. After the onset a postive-twisted flux rope with
dipped field lines forms during the eruption, which also pushes out
and reconnects with an outer negative-twisted field.
This interaction results in a complex erupting structure containing
two distinct flux ropes, and produces an outgoing double-shell
feature similar to that seen in the STEREO-B/EUVI 171 passpand
observation of the CME.
Furthermore, we find that the foot points of the erupting field lines
spatially agree with the major EUV dimming regions of the eruptive flare
observed by the SDO/AIA.
The change of the horizontal magnetic field at the lower boundary
during the simulated eruption also shows a pattern similar to
that of the observed change of the photosphere horizontal magnetic field
produced by the observed X-class flare.
These agreements suggest the validity of the modeled magnetic field
evolution for the initiation of this observed CME event.

The twisting electric field used here is a way to establish
the non-potential pre-eruption magnetic field in an accelerated
time-scale, capturing the cumulative effects of the long-term
build up (by e.g. shearing at the PIL and sunspot rotation)
as measured by the observed vertical electric current density.
It assumes an {\it ad hoc} constant transport speed $v_0$ that
is not well constrained by observations, and whose value
used in the current simulation is crudely selected by trial
and error with multiple simulations to best match the observed
eruptive behavior for the first eruption.
The choice of the value for $v_0$ is strongly influenced
by the numerical resistivity in the code in order to build up
the pre-eruption electric current against the numerical
dissipation, and the use of a constant value of $v_0$ continuously
results in the subsequent repeated eruptions which are not in
agreement with the observation.
Future improvements of the formulation of the twisting electric
field to temporally and spatially vary the transport 
speed based on additional observational constrains
are needed to improve the agreement of the modeled eruptive
behavior with the observations.
One improvement for example is to vary the
transport speed based on the difference between the simulated
and the observed vertical electric current at the lower boundary.


\acknowledgments
We thank the anonymous referee for helpful comments that improve the paper.
This material is based upon work supported by
the National Center for Atmospheric Research (NCAR),
which is a major facility sponsored by the National Science Foundation under Cooperative
Agreement No. 1852977. This work is also support by the NASA LWS grant 80NSSC19K0070
and the NASA HFORT grant 80NSSC22M0090's subaward to NCAR.
A.N.A.  acknowledges the support of NASA ECIP NNH18ZDA001N and DKIST Ambassador Program.
Funding for the DKIST Ambassadors program is provided by the National Solar Observatory,
a facility of the National Science Foundation, operated under Cooperative Support
Agreement number AST-1400450.
Resources supporting this work were provided by the NASA High-End Computing (HEC)
Program through the NASA Advanced Supercomputing (NAS) Division at Ames Research
Center. We also acknowledges high-performance computing support from Cheyenne
(doi:10.5065/D6RX99HX) provided by NCAR's Computational and Information Systems
Laboratory, sponsored by the National Science Foundation.


\end{document}